\documentclass[preprint,prd,noshowpacs]{revtex4}
\usepackage{amssymb}
\usepackage{amsmath}
\usepackage{graphicx}

\begin{document}

\title{Revisiting the Wu-Yang approach to magnetic charge}

\author{Siva Mythili Gonuguntla}
\email{sivamythili01@mail.fresnostate.edu}
\affiliation{Physics Department, California State University Fresno, Fresno, CA 93740 USA}

\author{Douglas Singleton}
\email{dougs@mail.fresnostate.edu}
\affiliation{Physics Department, California State University Fresno, Fresno, CA 93740 USA\\}
\affiliation{Kavli Institute for Theoretical Physics, University of California Santa Barbara, Santa Barbara, CA 93106, USA}

\date{\today}

\begin{abstract}
The Wu-Yang fiber bundle approach to magnetic charge is extended with a disk-like sheet current density and associated magnetic field in the overlap region between the Northern hemisphere and Southern hemisphere, where the different vector potentials connect. This disk magnetic field plays a role similar to the Dirac string in the Dirac approach to magnetic charge - it brings an inward magnetic flux of $4 \pi g$ which then gives rise to an outward Coulomb magnetic flux of $4 \pi g$. As with the Dirac string approach we show that placing an electric charge near this disk magnetic field gives rise to a non-zero electromagnetic field momentum.  We discuss some of the possible physical consequences of this electromagnetic field momentum. We conclude by showing that the non-singular, but non-single valued Banderet monopole potential also has a disk-like magnetic flux and non-zero electromagnetic field momentum in the presence of an electric charge.       
\end{abstract}

\maketitle

\section{Wu-Yang approach to magnetic charge}

A pure magnetic charge $g$ is defined as having a Coulomb magnetic field ${\bf B} = \frac{g {\bf r}}{r^3}$ which implies $\nabla \cdot {\bf B} = 4 \pi g \delta ({\bf r})$. This last equation runs afoul of the relationship between the magnetic field and the vector potential namely ${\bf B} = \nabla \times {\bf A}$. For a well behaved ${\bf A}$ one has $\nabla \cdot {\bf B} = \nabla \cdot (\nabla \times {\bf A}) = 0 \ne 4 \pi g \delta ({\bf r})$. 

In the Dirac string approach to magnetic charge  the string potentials are \cite{dirac,dirac1}
\begin{equation}
\label{A-coulomb}
    {\bf A} _\pm ({\bf x}) = \frac{g}{r} \left( \frac{\pm 1 - \cos \theta }{\sin \theta} \right) {\bf{\hat \varphi}} ~ =\frac{g}{\rho} \left( \pm 1 - \frac{z}{\sqrt{\rho^2+z^2}} \right) {\bf{\hat \varphi}} ~.
\end{equation}
Taking the curl of \eqref{A-coulomb} gives a Coulomb magnetic field $\nabla \times {\bf A} _\pm ({\bf x}) = \frac{g {\bf r}}{r^3}$.  ${\bf A} _\pm ({\bf x})$ is given in both spherical and cylindrical coordinates since later when we look at the Wu-Yang approach to magnetic charge the cylindrical version will be the most useful. 

From \eqref{A-coulomb} the two vectors potentials are not well behaved since they are singular along the entire negative/positive z-axis for ${\bf A} _+ (r)/{\bf A} _- (r)$.  Wu and Yang dealt with these string singularities using a fiber bundle approach \cite{wu-yang,yang}. They took ${\bf A}_+$ as the vector potential over the Northern hemisphere and ${\bf A}_-$ as the vector potential over the Southern hemisphere. This avoided the string singularity in each region. One can write the extended Wu-Yang fiber bundle potential mathematically as
\begin{equation}
\label{fiber}
    {\bf A} _{WY} ({\bf x})  =\frac{g}{\rho} \left( + \Theta (+ z +\epsilon) - \Theta (-z +\epsilon)- \frac{z}{\sqrt{\rho^2+z^2}} \right) {\bf{\hat \varphi}} ~.
\end{equation}
$\Theta (x)$ is the step function, which equals $1$ when the argument is positive and equals $0$ when the argument is negative. Note that $0< \epsilon \ll 1$ is a small, positive constant. This parameter $\epsilon$ does not need to be infinitesimal. In fact the limit $\epsilon \to 0$ can not be reached since the cover sets of the two sphere ({\it i.e.} $S^2$) need to be open sets in fiber bundle theory. The $\Theta$ functions in \eqref{fiber} are a concrete way (but not necessarily the only way) of realizing the Wu-Yang fiber bundle construction of the vector potential. The vector potential in \eqref{fiber} has replaced the $\pm 1$ of \eqref{A-coulomb} with $+ \Theta (+ z +\epsilon) - \Theta (-z +\epsilon)$, and it is this which will lead to an additional disk-like singularity in the magnetic field, as we will see shortly.

The two potentials, ${\bf A} _+ ({\bf x})$ and ${\bf A} _- ({\bf x})$, are related by a gauge transforms of the form 
\begin{equation}
\label{A-gauge}
{\bf A} _+ ({\bf x})-{\bf A} _- ({\bf x}) = \nabla _{\varphi} \alpha = \frac{2g}{\rho} {\hat \varphi} ~.
\end{equation}
The gauge function $\alpha = 2 g \varphi$ is non-single valued. However as pointed out in section II of reference \cite{wu-yang} this can be dealt with using the fiber bundle approach. 

The Dirac quantization condition is obtained by placing an electric charge, $q$, in the neighborhood of the Wu-Yang monopole and noting that the wavefunction of $q$ in the monopole background will be broken up into a Northern and Southern hemisphere wavefunction - $\Psi _+ ({\bf x})$ and $\Psi _- ({\bf x})$ respectively. These wave functions are related by the standard wavefunction gauge transformation \cite{wu-yang,yang}  
\begin{equation}
\label{wave-gauge}
\Psi _+ ({\bf x})= e^{iq \alpha / \hbar} \Psi _- ({\bf x}) = e^{2iqg\varphi /\hbar} \Psi_- 
\end{equation}
If the wavefunctions $\Psi _+ ({\bf x})$ and $\Psi _- ({\bf x})$ are to match up as $\varphi$ goes from $0$ to $2 \pi$ one needs $\frac{2qg}{\hbar}=n$, where $n$ is an integer. This ensures that the factor $e^{2iqg\varphi /\hbar}=1$ and $\Psi _+ ({\bf x})$ and $\Psi _- ({\bf x})$, from \eqref{wave-gauge}, will match. The condition $\frac{2qg}{\hbar}=n \to qg = n \frac{\hbar}{2}$ is the Dirac quantization condition, obtained from the Wu-Yang approach. 

While the wavefunctions in the Northern and Southern hemispheres now match at the equator, equations \eqref{A-coulomb} and \eqref{fiber} show that there is a jump in the vector potential in going from the Northern hemisphere, with ${\bf A}_+$, to the Southern hemisphere, with ${\bf A}_-$. This jump in the vector potential results in a previously overlooked ``disk" magnetic field in the equatorial overlap region. This disk magnetic field is similar to the Dirac string, but instead of being confined to a string singularity, there is a disk discontinuity. 

The magnetic field in the Wu-Yang approach is obtained by taking the curl of the Wu-Yang vector potential in \eqref{fiber} giving 
\begin{equation}
    \label{curl-wy}
    {\bf B}= \nabla \times {\bf A} _{WY} = -\partial _z (A^\varphi _{WY} ) {\bf \hat \rho} + \frac{1}{\rho} \partial _{\rho} (\rho A^{\varphi} _{WY}) {\bf \hat z}~.
\end{equation}
Plugging the $- \frac{g~z}{\rho \sqrt{\rho^2+z^2}}{\bf{\hat \varphi}}$ term from \eqref{fiber} into \eqref{curl-wy} gives $g \frac{\rho {\hat {\bf \rho}} + z {\hat {\bf z}}}{(\rho ^2 +z^2)^{3/2}}$ which is the Coulomb magnetic field in cylindrical coordinates. At first it might seem that this would be all to the curl of the Wu-Yang vector potential, but the $+ \frac{g \Theta (+ z + \epsilon)}{\rho}$ and $- \frac{g \Theta (- z+ \epsilon)}{\rho}$ terms of ${\bf A}_{WY} ({\bf x})$ yield a non-zero result since $ {\lim}_{\epsilon \to 0} ~ \frac{d}{dz} \left(+ \Theta (+ z + \epsilon) \right) = \delta (z)$ and ${\lim}_{\epsilon \to 0} ~\frac{d}{dz}  \left(- \Theta (- z + \epsilon) \right) = \delta (-z) = \delta (z)$ (the last step used the fact that the delta function is a even function). Putting this together gives a magnetic field of the form
\begin{equation}
    \label{b-disk}
{\bf B} = g \frac{\rho {\hat {\bf \rho}} + z {\hat {\bf z}}}{(\rho ^2 +z^2)^{3/2}} - \frac{2g \delta (z)}{\rho} {\hat {\bf \rho}}~.
\end{equation}
This second term is \eqref{b-disk} has to the best of our knowledge been overlooked in previous work on the Wu-Yang monopole.  In a comment on an earlier version of this work \cite{comment} the authors criticize the use of the vector potential in \eqref{fiber} by simply stating that this does not conform to the fiber bundle approach of the original work by Wu and Yang \cite{wu-yang}. In this revised version we address the points raised in \cite{comment}. We emphasize that we do not require the overlap region to be taken in the zero width limit. Also in the added next section we obtain the  delta function term, without explicitly using the vector potential in \eqref{fiber}, but by using the boundary conditions as one crosses the region where ${\bf A}_+$ and ${\bf A}_-$ overlap. The authors of the comment say that ``We did not check the details of these calculations." To address this admission of not `` checking the calculations" we have provided calculational details so that our claims can be explicitly and easily checked.    

\section{Disk magnetic field from boundary condition} 

Because the result in \eqref{b-disk} is central to our analysis (in particular the existence of the  delta function term) we want to obtain \eqref{b-disk} without using the explicit form of the vector potential from \eqref{fiber}, but rather using boundary conditions. We will follow closely the analysis of the original Wu -Yang work \cite{wu-yang}, including the figures, to address the issues raised in \cite{comment}.  To check for the delta function term in \eqref{b-disk} we will use the integral relationship between ${\bf B}$  and ${\bf A}$ ({\it i.e.} $\int {\bf B} \cdot d {\bf a} = \oint {\bf A} \cdot d {\bf l}$) for the contour shown in Fig. 1.
\begin{figure}[ht]
 \centering
\includegraphics[width=80mm]{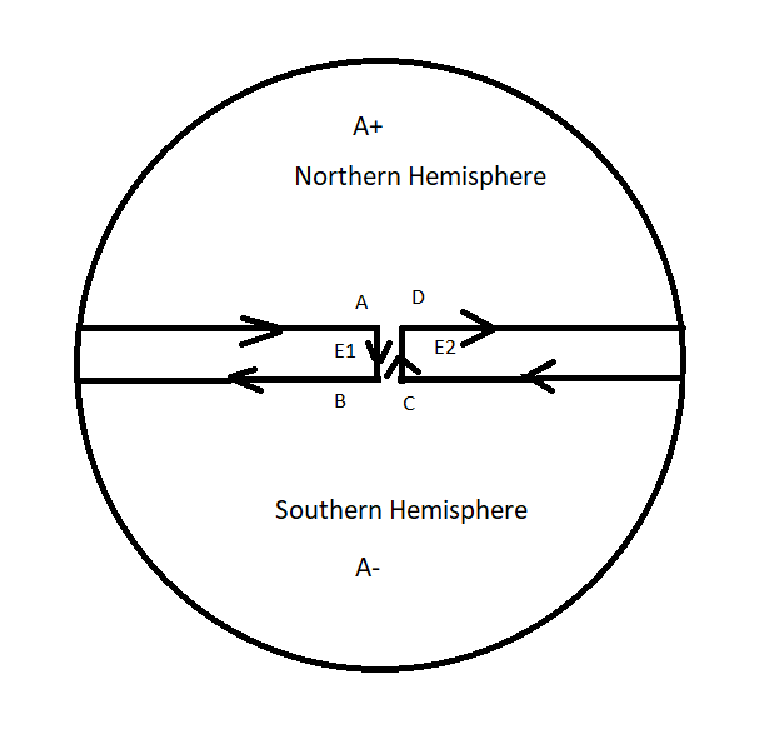}
 \caption{The contour for evaluating $\int {\bf B} \cdot d {\bf a} = \oint {\bf A} \cdot d {\bf l}$ across the equator.}
\label{fig1}
\end{figure}
The upper part of the contour in Fig. 1, goes left to right from point $D$ to point $A$. This part of the contour lies entirely in the Northern hemisphere where the vector potential is purely ${\bf A}_+$. The lower part of the contour in Fig. 1. goes right to left from point $B$ to point $C$. This part of the contour lies entirely in the Southern hemisphere where the vector potential is purely ${\bf A}_-$. Finally, the two parts of the contour labeled $E1$ and $E2$ cross the equator going from North to South, for $E1$, and from South to North for $E2$. These parts of the contour which cross the equator are effectively``on top of one another".   

Fig. 2 shows more detail about the path in Fig. 1. The dashed lines above and below the equator mark the point above which/below which the vector potential becomes purely ${\bf A}_+$/${\bf A}_-$. As per equation \eqref{fiber} the upper/lower dashed line in Fig. 2 is $\epsilon$ above/below the equator.  
\begin{figure}[ht]
 \centering
\includegraphics[width=120mm]{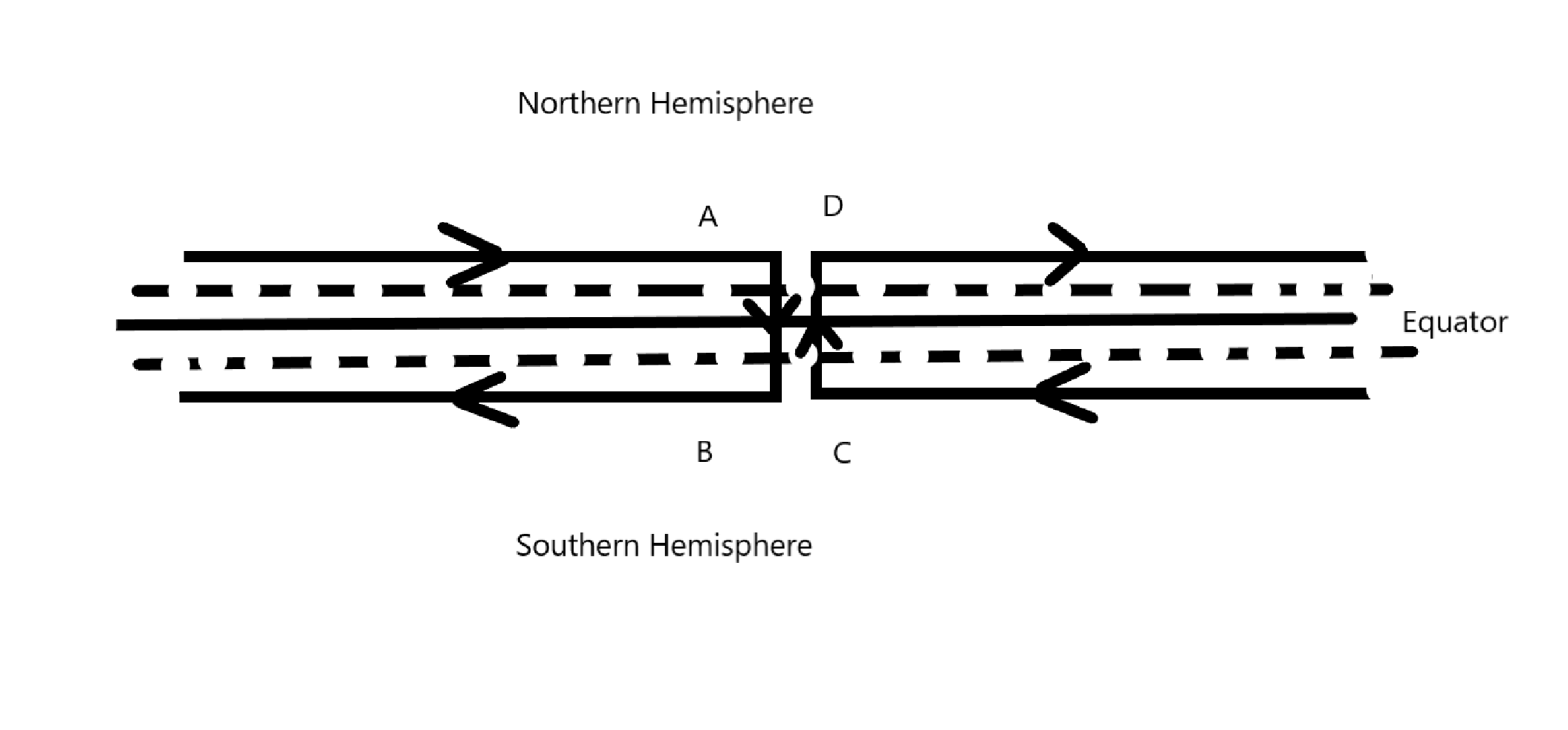}
 \caption{Details of the path used to evaluate $\int {\bf B} \cdot d {\bf a} = \oint {\bf A} \cdot d{\bf l}$ for the path in Fig. 1. The upper/lower dashed lines are $\epsilon$ above/below the equator and are the boundary above which/below which the vector potential is purely ${\bf A}_+$/${\bf A}_-$. The solid lines are the contour of the line integral from $\oint {\bf A} \cdot d {\bf l}$. The upper path from $D$ to $A$ is $\epsilon '$ above the equator, and the lower path from $B$ to $C$ is $\epsilon '$ below the equator. These distances are related by $0< \epsilon < \epsilon ' \ll 1$.}
\end{figure}
The parts of the path in Figs. 1 and 2, that go from $D$ to $A$ and from $B$ to $C$, are straight forward to evaluate since they are in regions where the vector potential is entirely ${\bf A}_+$ or ${\bf A}_-$. The path going from $D$ to $A$ gives
\begin{eqnarray}
\label{DA}
    \int _D ^A {\bf A}_+ \cdot d {\bf l} &=& \int _0 ^{2 \pi} \frac{g}{\rho _0} \left( 1- \frac{\epsilon '}{\sqrt{\rho_0 ^2 + (\epsilon')^2}}\right) {\hat \varphi} \cdot \rho_0 d \varphi {\hat \varphi} \nonumber \\
    &=& 2 \pi g -\frac{2 \pi g \epsilon '}{\sqrt{\rho_0 ^2 + (\epsilon')^2}}~.
\end{eqnarray}
The path going from $B$ to $C$ gives
\begin{eqnarray}
\label{BC}
    \int _B ^C {\bf A}_- \cdot d {\bf l} &=& \int _0 ^{2 \pi} \frac{g}{\rho _0} \left( -1- \frac{(-\epsilon ')}{\sqrt{\rho_0 ^2 + (-\epsilon')^2}}\right) {\hat \varphi} \cdot (-\rho_0 d \varphi {\hat \varphi}) \nonumber \\
    &=& 2 \pi g -\frac{2 \pi g \epsilon '}{\sqrt{\rho_0 ^2 + (\epsilon')^2}}~,
\end{eqnarray}
To make a closed loop we need to add in the short segments $A$ to $B$ and $C$ to $D$. Naively these line integrals appear to be zero ({\it i.e.} $\int _A ^B {\bf A}_{\pm} \cdot d{\bf l} = \int _C ^D {\bf A}_{\pm} \cdot d{\bf l} = 0$) since ${\bf A}_{\pm} \propto {\hat \varphi}$ while $d{\bf l} \propto {\hat {\bf z}}$ (if one uses cylindrical coordinates) or $d{\bf l} \propto {\hat {\bf \theta}}$ (if one uses spherical coordinates) and in both cases ${\bf A}_{\pm} \cdot d{\bf l} =0$.

However, section I of reference \cite{wu-yang} emphasized that care should be taken for paths that cross from one fiber to the other as is the case for the line segments $A$ to $B$ and $C$ to $D$. For such segments it is the exponential of the loop integral ( {\it i.e.} $\exp \left[ \frac{iq}{\hbar }  \oint A_\mu dx^\mu \right]$, equation 2 of \cite{wu-yang}) which is physically meaningful rather than the loop integral by itself ({\it i.e.}  $\frac{q}{\hbar}  \oint A_\mu dx^\mu $, equation 1 of \cite{wu-yang}). Note here we use $q$ for the electric charge versus reference \cite{wu-yang}, which uses $e$, and we are setting $c=1$ throughout the paper. Thus for the short segments $A$ to $B$ and $C$ to $D$ we want to look at $\Phi _{AB} = \exp \left ( \frac{iq}{\hbar } \int ^B _A {\bf A} \cdot d{\bf l} \right)$ and $\Phi _{CD} = \exp \left ( \frac{iq}{\hbar } \int ^D _C {\bf A} \cdot d{\bf l} \right)$ respectively. 

For both of the line integrals, $\int ^B _A {\bf A} \cdot d{\bf l}$ and $\int ^D _C {\bf A} \cdot d{\bf l}$, one needs to know which vector potential to use - ${\bf A}_+$ or ${\bf A}_-$ - as one crosses the equator. Wu and Yang showed that this connection is accomplished with the gauge transformations of the form $S_{EA-EB} = \exp \left( \frac{2iqg}{\hbar } \varphi \right)$. With this gauge transformation $\Phi _{AB}$ is given by 
\begin{eqnarray}
    \label{AB}
    \Phi _{AB} &=& \Phi _{A-EA} S_{EA-EB} \Phi _{B-EB} \nonumber  \\
    &=& \exp \left ( \frac{iq}{\hbar } \int ^{EA} _A {\bf A}_+ \cdot d{\bf l} \right)  \exp \left( \frac{2iqg}{\hbar } \varphi_0 \right) \exp \left ( \frac{iq}{\hbar } \int ^B _{EB} {\bf A}_- \cdot d{\bf l} \right) \\
    &=& \exp (0) \exp \left( \frac{2iqg}{\hbar } \varphi_0 \right)  \exp (0) = \exp \left( \frac{2iqg}{\hbar } \varphi_0 \right) ~,\nonumber
\end{eqnarray}
where $\varphi_0$ is the azimuthal angle of the path $A \to EA \to EB \to B$. The details of the calculation in \eqref{AB} are shown graphically in Fig. 3 which is similar to Fig. 2 of reference \cite{wu-yang}. The path integrals $\int ^{EA} _A {\bf A}_+ \cdot d{\bf l}$ and $\int ^B _{EB} {\bf A}_- \cdot d{\bf l}$ are both zero since for these paths ${\bf A}_\pm \propto {\hat \varphi}$ while $d{\bf l} \propto dz {\hat {\bf z}}$, when using cylindrical coordinates, so ${\bf A}_\pm \cdot d {\bf l} =0$.  

\begin{figure}[ht]
 \centering
\includegraphics[width=100mm]{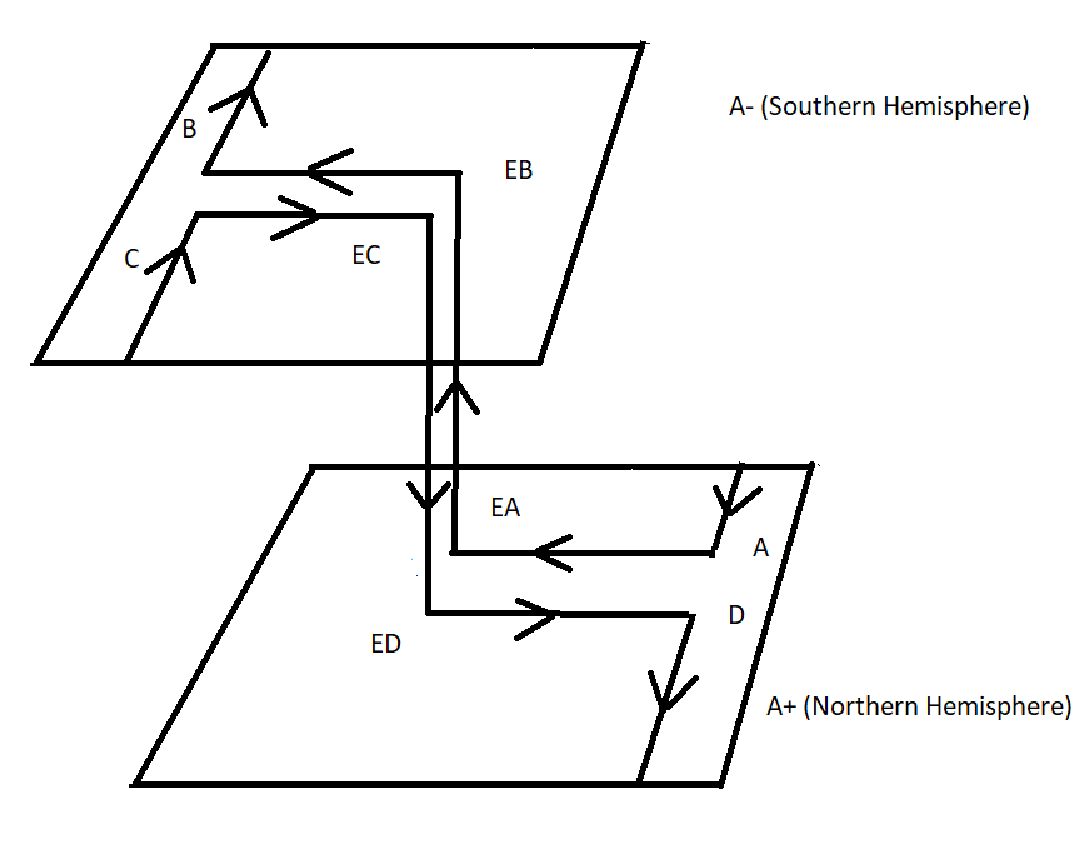}
 \caption{Details of crossing the equator for the contour in Fig. 1.}
\label{fig3}
\end{figure}

After reaching $B$ one takes a path all the way around the sphere (going  from $\varphi_0$ to $\varphi_0 - 2 \pi$) to point $C$. This path integral is entirely in the Southern hemisphere and the result is given in \eqref{BC}. After reaching point $C$ we again need to cross the equator which gives 
\begin{eqnarray}
    \label{CD}
    \Phi _{CD} &=& \Phi _{C-EC} S_{EC-ED} \Phi _{ED-D} \nonumber  \\
    &=& \exp \left ( \frac{iq}{\hbar } \int ^{EC} _C {\bf A}_- \cdot d{\bf l} \right)  \exp \left( - \frac{2iqg}{\hbar } (\varphi_0 + 2 \pi) \right) \exp \left ( \frac{iq}{\hbar } \int ^D _{ED} {\bf A}_+ \cdot d{\bf l} \right) \\
    &=& \exp (0) \exp \left( -\frac{2iqg}{\hbar } (\varphi_0 - 2 \pi) \right)  \exp (0) = \exp \left( - \frac{2iqg}{\hbar } (\varphi_0 - 2 \pi) \right) \nonumber ~,
\end{eqnarray}
where $\varphi_0 - 2 \pi$ is the azimuthal angle of the path $C \to EC \to ED \to D$. This is shown in Fig. 3. The path integrals $\int ^{EC} _C {\bf A}_- d{\bf l}$ and $\int ^D _{ED} {\bf A}_+ d{\bf l}$ are both zero since for these paths  ${\bf A}_\pm \propto {\hat \varphi}$ while $d{\bf l} \propto dz {\hat {\bf z}}$, so ${\bf A}_\pm \cdot d{\bf l} \propto {\hat \varphi} \cdot {\hat {\bf z}} =0$.

Using the results from \eqref{DA} \eqref{BC} \eqref{AB} and \eqref{CD} we can now obtain the correct closed loop line integral going from $D \to A \to B \to C \to D$. The parts of the line integral from $D \to A$ and $B \to C$ are simple since these parts of the path are in regions where the vector potential is either ${\bf A}_+$ or ${\bf A}_-$. These results are given in \eqref{DA} and \eqref{BC} respectively. The parts of the path that cross the equator - $A \to B$ or $C \to D$ - need to be handled via the $\Phi_{AB}$ and $\Phi _{CD}$ given in \eqref{AB} and \eqref{CD} which involve the use of the gauge factor $S_{EA-EB}$ and $S_{EC-ED}$. These factors combine to give
\begin{eqnarray}
    \label{ABCD-2}
    \Phi _{AB} \Phi _{CD} &=& 
    = \exp \left( \frac{2iqg}{\hbar } \varphi_0 \right)  \exp \left( - \frac{2iqg}{\hbar } (\varphi_0 - 2 \pi) \right)  \\
    &=&  \exp \left( \frac{2iqg}{\hbar } (2 \pi) \right) = 1~. \nonumber 
\end{eqnarray}
In the second line of \eqref{ABCD-2} use have used the Dirac quantization condition, namely $\frac{2qg}{\hbar } =n$, which immediately gives $\exp(i 2 \pi n) =1$. Since the factor in \eqref{ABCD-2} is 1, the line integrals inside these exponents total to zero. Taking all of these results from \eqref{DA} and \eqref{BC} and the implication from \eqref{ABCD-2} together, the closed loop integral from Figs. 1, 2 and 3 is
\begin{eqnarray}
    \label{ABCD-3}
    \oint {\bf A} \cdot d{\bf l} &=& 
    \int _D ^A {\bf A}_+ \cdot d{\bf l} + \int_A ^B {\bf A} \cdot d{\bf l} + \int _B ^C {\bf A}_- \cdot d {\bf l} + \int _C ^D {\bf A} \cdot d {\bf l}   \nonumber \\
    &=& \left( 2 \pi g -\frac{2 \pi g \epsilon '}{\sqrt{\rho_0 ^2 + (\epsilon')^2}}\right) + 0 + \left(2 \pi g -\frac{2 \pi g \epsilon '}{\sqrt{\rho_0 ^2 + (\epsilon')^2}}\right) + 0 ~ \\
    &=& 4 \pi g - \frac{4 \pi g \epsilon '}{\sqrt{\rho_0 ^2 + (\epsilon')^2}} \nonumber 
\end{eqnarray}
By Stokes' theorem the result in \eqref{ABCD-3} should be equal to the surface area integral $\int {\bf B} \cdot d{\bf a}$, where the surface is bounded by the closed contour. There are two possible areas one could use for the closed line integral shown in Fig. 1: (i)  One could take the area as the entire sphere minus the thin strip around the equator, or (ii) one could take the area as just thin strip around the equator.

For the choice of area (i) looking at the result in \eqref{ABCD-3} one sees that the $4 \pi g$ is just the magnetic flux of a magnetic monopole $g$ through the entire sphere and the term $- \frac{4 \pi g \epsilon '}{\sqrt{\rho_0 ^2 + (\epsilon')^2}}$ is just the loss flux of magnetic flux due to the exclusion of the thin trip around the equator. Very briefly for this strip ${\bf B} \approx  \frac{g}{(\rho_0 ^2 +(\epsilon ')^2)} {\hat \rho}$ and $\Delta {\bf a} \approx 2 \pi \sqrt{\rho_0 ^2 +(\epsilon')^2} \times 2 \epsilon ' {\bf \hat \rho}$. Thus the flux through the strip at the equator is ${\bf B} \cdot \Delta {\bf a} \approx  \frac{4 \pi g \epsilon '}{\sqrt{\rho_0 ^2 + (\epsilon')^2}}$. The minus sign in  $- \frac{4 \pi g \epsilon '}{\sqrt{\rho_0 ^2 + (\epsilon')^2}}$ comes since this flux is subtracted/excluded from the area bounded by the path in Fig. 1. For this area - the entire sphere minus equatorial strip - one finds that $\oint {\bf A} \cdot d {\bf l}$ from \eqref{ABCD-3} does equal $\int {\bf B} \cdot d {\bf a}$ so Stokes' theorem is satisfied. 

What about area choice (ii) - the thin strip around the equator? In this case the flux due to the monopole field at the origin is determined by ${\bf B} \approx  \frac{g}{(\rho_0 ^2 +(\epsilon ')^2)} {\hat \rho}$ and $\Delta {\bf a} \approx - 2 \pi \sqrt{\rho_0 ^2 +(\epsilon')^2} \times 2 \epsilon ' {\bf \hat \rho}$, which yields ${\bf B} \cdot \Delta {\bf a} \approx - \frac{4 \pi g \epsilon '}{\sqrt{\rho_0 ^2 + (\epsilon')^2}}$. Here the minus sign comes about due to the reversal of the vector area of the thin equatorial strip, relative to the area for case (i). We can now see that the delta function term ($- \frac{2g\delta (z)}{\rho} {\hat \rho}$ from \eqref{b-disk}) is needed to give a flux of $4 \pi g$ through the thin strip as is required by Stokes' theorem. Briefly for this thin strip the delta function term yields $\int {\bf B} \cdot d{\bf a} \approx \int \left( - \frac{2g \delta (z)}{\rho_0}\right) \cdot \left(-2 \pi \rho_0 \Delta z \right) = 4 \pi g$, where we have used the property of the delta function $\int \delta (z) \Delta z =1$. Thus $\oint {\bf A} \cdot d {\bf l} = \int {\bf B} \cdot d{\bf a}$ for both areas (i) and (ii) bounded by the contour in Fig. 1, and Stokes' theorem is satisfied {\it only if} one includes the delta function term in \eqref{b-disk}. Thus both by the explcit construction of section I and through the use of boundary conditions in this section, we have shown the need for the disk-like delta function term of \eqref{b-disk}.
 
Finally we note that the magnetic field from \eqref{b-disk} implies a surface current density via $\nabla \times {\bf B} = 4 \pi {\bf J}$. The first, Coulomb term in \eqref{b-disk} has a curl of zero since it is radially symmetric. The second, disk term gives
\begin{equation}
    \label{sheet}
    \nabla \times \left( - \frac{2g \delta (z)}{\rho} {\hat {\bf \rho}} \right) = -\frac{2g}{\rho} \frac{d(\delta (z))}{dz} {\hat {\bf \varphi}} = -\frac{2g}{\rho} \delta ' (z) {\hat {\bf \varphi}}~,
\end{equation}
which implies a surface current density of ${\bf J} = -\frac{g}{2 \pi \rho}  \delta ' (z) {\hat {\bf \varphi}}$. The fact that the current density is proportional to the derivative of a delta function implies that there is a {\it dipole} current density at $z=0$. The above analysis of the jump in the vector potential over a short distance and with a source that is the derivative of delta function, is essentially the magnetic version of the charge dipole layer jump of the scalar potential discussed in section 1.6 of Jackson's electrodynamics text \cite{jackson}.

\section{Comparison to Dirac string formulation}

In this section we compare the Wu-Yang approach to magnetic charge with the Dirac approach. In particular we show that the disk magnetic field of equation \eqref{b-disk} plays the same role as the Dirac string - it carries an inward flux of $4 \pi g$ which then emerges from the origin as a Coulomb magnetic field of outward flux $4 \pi g$. 

The Dirac approach to magnetic charge also leads to a delta function contribution to the ${\bf B}$ field, but one that is only along the $z$-axis rather than in the entire $z$-plane as in \eqref{b-disk}. To deal with the string singularity of the vector potential \eqref{A-coulomb} one defines a regularized vector potential as ${\bf A}^{regular}_{\pm} = \frac{g \Theta (\rho -\epsilon)}{\rho} \left( \pm 1 - \frac{z}{\sqrt{\rho^2+z^2 + \epsilon^2}} \right) {\bf{\hat \varphi}}$ (see appendix D of \cite{heras} for details of this calculation). Taking the curl of ${\bf A}^{regular}_{\pm}$ and the limit $\epsilon \to 0$ gives
\begin{eqnarray}
\label{b-coulomb}
   {\bf B} &=&  \lim_{\epsilon \to 0} \nabla \times ({\bf A}^{regular} _\pm ) = g \frac{\rho {\hat {\bf \rho}} + z {\hat {\bf z}}}{(\rho ^2 +z^2)^{3/2}}\pm 2  g \frac{\delta (\rho)}{\rho} \Theta (\mp z) {\bf {\hat z}} \nonumber \\
   &=& g \frac{{\hat {\bf r}}}{r^2} \pm 4 \pi g \delta (x) \delta (y) \Theta (\mp z) {\hat {\bf z}} ~,
\end{eqnarray}
where we used $\frac{\delta (\rho)}{2 \pi \rho} = \delta (x) \delta (y)$. This form of the magnetic field is explicitly derived and discussed in several review articles and monographs \cite{heras,olive,blag,felsager,shnir,adorno,mavromatos}.
The first term on the right hand side of \eqref{b-coulomb} is the point, Coulomb part and the second term is the delta, string contribution. Requiring that this string magnetic field in \eqref{b-coulomb} have no observable effect on an electric charge $q$, leads to the Dirac quantization condition $qg= n \frac{\hbar}{2}$
\cite{heras,olive,blag,felsager,shnir,adorno,mavromatos}.

The delta function terms of the Wu-Yang and Dirac approaches - equations \eqref{b-disk} and \eqref{b-coulomb} respectively - both carry an inward flux of $-4 \pi g$ toward the origin, which then gives rise the outward flux of $+4 \pi g$ coming from the Coulomb magnetic field terms in \eqref{b-disk} and \eqref{b-coulomb}. This can be seen by taking a surface integral for both delta function terms in \eqref{b-disk} and \eqref{b-coulomb} {\it i.e.} $\oint {\bf B}_{\delta-term} \cdot d{\bf a}$. The surface we take is a cylindrical ``box" of height $2H$ and radius $\rho = R$ centered at the origin. The sides of the this surface have an integration of $\int_{-H} ^{+H} \int _0 ^{2 \pi} (...) \cdot {\hat {\bf \rho}} R d \varphi dz$ and the top/bottom have surface integrals $\int_0 ^R \int _0 ^{2 \pi} (...) \cdot ( \pm {\hat {\bf z}}) \rho d \rho d \varphi$ where $+(-)$ is for the top (bottom) surface. For the integral of the Wu-Yang delta term from \eqref{b-disk} this yields
\begin{equation}
    \label{d-disk}
    - \oint \left( \frac{2g \delta (z)}{\rho} {\hat \rho} \right) \cdot d{\bf a} = -2 \pi \int _{-H} ^{+H} \delta (z) \frac{2g}{R} {\hat \rho}\cdot {\hat \rho} R dz = - 4 \pi g ~. 
\end{equation}
Only the curved sides of the cylindrical box  contribute, the $d \varphi$ integration gives $2 \pi$, and the $dz$ integration gives unity, yielding an inward flux of $4 \pi g$. Using the same surface to integrate the Dirac string delta function term in \eqref{b-coulomb} gives
\begin{equation}
    \label{d-coulomb}
    \mp \oint \left( \frac{2 g \delta (\rho)}{\rho} \Theta (\pm z) {\bf {\hat z}} \right) \cdot d{\bf a} = \mp 2 \pi \int _0 ^R \frac{2g \delta (\rho)}{\rho} {\hat z}\cdot ({\pm \hat z}) \rho d \rho = - 4 \pi g ~. 
\end{equation}
Now only the top/bottom of the cylindrical box contributes, the $d \varphi$ integration again gives $2 \pi$ and the $d \rho$ integration gives unity, yielding an inward flux of $4 \pi g$. This inward flux in \eqref{d-disk} and \eqref{d-coulomb} is balanced by an outward flux of $4 \pi g$ coming from the Coulomb part of the magnetic field. 

The above analysis, showing the need for the delta function term in \eqref{b-disk} based on calculations of magnetic flux, supports the results of section II which arrived at the need for the delta function term by looking at boundary conditions across the equator.

\section{Electromagnetic Field Momentum}

We now show that the delta function term from the magnetic field in \eqref{b-disk} leads to a non-vanishing electromagnetic field momentum in the presence of an electric charge. To salvage momentum conservation there should be some ``hidden" momentum \cite{coleman,boyer,griffiths,griffiths-2} coming from the charges/currents in the system to balance the field momentum. A similar balancing of field momentum and ``hidden" momentum was shown for the Dirac string approach to magnetic charge in \cite{siva}.

The electromagnetic field momentum of an electric charge, $q$, plus the Wu-Yang monopole \eqref{b-disk} is obtained by integrating ${\bf E} \times {\bf B}^{disk}$. The Coulomb part of the Wu-Yang monopole's magnetic field would contribute a term of the form ${\bf E} \times \frac{g {\bf r}}{r^3}$, but by symmetry this term leads to zero field momentum. Using cylindrical symmetry, and without loss of generality, we place the electric charge, $q$ at ${\bf r}_0 = \rho _0 {\bf \hat x} + z_0 {\bf \hat z}$ {\it i.e.} along the $x$-axis. The electric field is then given by ${\bf E} = q \frac{{\bf r} '}{r'^3}$ where ${\bf r}' = {\bf r} - {\bf r}_0$. In cylindrical coordinates  $r' = (\rho ^2+ \rho_0 ^2 - 2 \rho \rho_0 \cos \varphi+ (z-z_0)^2)^{1/2}$. Putting this all together gives   
\begin{eqnarray}
\label{momentum}
{\bf P}_{EM} ^{disk} &=& \frac{1}{4 \pi} \int q  \frac{{\bf \hat r}'}{{r'}^2} \times \left( - \frac{2 g \delta (z)}{\rho} {\bf {\hat \rho}} \right) d^3 x \nonumber \\
&=& - \frac{qg}{2 \pi} \int _0 ^\infty \rho d \rho \int _0 ^{2 \pi} d \varphi  \frac{\left[ \rho {\bf \hat \rho} - \rho_0 {\bf \hat x} -z_0 {\bf \hat z} \right]}{((\rho ^2 + \rho_0 ^2 + z_0 ^2 -2 \rho \rho_0 \cos \varphi)^{3/2}}\times \frac{{\bf \hat \rho}}{\rho}  \\
&=&-\frac{qg}{2 \pi} \int _0 ^\infty  d \rho \int _0 ^{2 \pi} d \varphi  \left( \frac{z_0 \sin \varphi {\bf \hat x} - z_0 \cos \varphi {\bf \hat y} -\rho_0 \sin \varphi {\bf \hat z} }{(\rho ^2 + \rho_0 ^2 + z_0 ^2 -2 \rho \rho_0 \cos \varphi)^{3/2}} \right) \nonumber ~.
\end{eqnarray} 
In going from the first to second line we have done the $dz$ integration using the delta function. The $d \varphi$ integration of the ${\bf \hat x}$ and ${\bf \hat z}$ components in the third line are of the same form and both give zero. The integration of the ${\bf \hat y}$ component gives
\begin{equation}
    \label{momentum-2}
    {\bf P}_{EM} ^{disk} = - \frac{qg z_0}{\rho _0 r_0} {\bf \hat y}
\end{equation}
When the electric charge is on the $z$-axis one needs to go back to  the integral in \eqref{momentum} and set $\rho_0 = 0$. The $d \varphi$ integration then gives ${\bf P}^{disk}_{EM}|_{\rho_0 =0} =0$. The field momentum is zero both when the electric charge is in the $z$-plane ($z_0 =0$) and when the electric charge is on the $z$-axis, but otherwise is non-zero. This is a violation of the center of energy theorem \cite{zangwill,coleman} since neither the electric charge nor the magnetic charge are moving and yet there appears to be some non-zero momentum in the electromagnetic field given by \eqref{momentum-2}.

This type of paradox -- a system whose parts are not moving and yet has some non-zero momentum in the fields -- was pointed out by Shockley and James in \cite{shockley} and the resolution in terms of ``hidden momentum" (momentum carried by the charged particles responsible for the currents in the system) was given in \cite{coleman}. This is a subtle issue that has been discussed in several excellent pedagogical articles \cite{boyer,griffiths,griffiths-2}. 

The existence of the non-zero field momentum in \eqref{momentum-2} is problematic for the Wu-Yang formulation of magnetic charge since either: (i) ${\bf P}_{EM} ^{disk} \ne 0$ implying a violation of the center of energy theorem \cite{zangwill,coleman}, or (ii) there is some ``hidden" momentum in the system to balance the non-zero field momentum. However having a ``hidden" momentum implies there are additional charge/current densities for this system, and thus the system is no longer a pure magnetic charge. 

Let us examine the second option that there is some ``hidden" mechanical momentum to balance the electromagnetic field momentum from \eqref{momentum-2}. The mechanical momentum stored in charges and currents is given by \cite{griffiths,griffiths-2}
 \begin{equation}
     \label{hid-mom}
     {\bf P}_{mech} ^{hid} = - \int \phi {\bf J} d^3x~.
 \end{equation}
The current density in \eqref{hid-mom} is given below \eqref{sheet} as ${\bf J} = -\frac{g}{2 \pi \rho}  \delta ' (z) {\hat {\bf \varphi}}$, and the potential for the charge $q$ in \eqref{hid-mom} is given by  $\phi =\frac{q}{\sqrt{(x-x_0)^2 +y^2+(z-z_0)^2}}$. Note that without loss of generality, due to the cylindrical symmetry of the magnetic field, we have placed the charge $q$ at  ${\bf r}_0 = (x_0, 0, z_0)$. At the end $x_0$ is equivalent to $\rho_0$, the cylindrical radial distance. 
 \begin{equation}
     \label{hid-mom-2}
     {\bf P}_{mech} ^{hid} = \frac{gq}{2 \pi} \int  \frac{\delta '(z) (-\sin \varphi {\bf \hat x} + \cos \varphi {\bf \hat y})}{\rho \sqrt{\rho ^2+ x_0^2 +(z-z_0)^2 -2 x_0 \rho \cos \varphi }} d^3x~.
 \end{equation}
 For the ${\bf \hat x}$-term in \eqref{hid-mom-2} the $d \varphi$ integration is of the form $\int _0 ^{2 \pi} \frac{\sin \varphi }{\sqrt{A-B \cos \varphi}} d \varphi$, which equals zero. Thus there is no ${\bf \hat x}$ component to the hidden mechanical momentum which is consistent with \eqref{momentum-2}. For the ${\bf \hat y}$-term we first do the $dz$-integration via an integration by parts
 \begin{eqnarray}
 \label{dz}
 &&\frac{gq}{2 \pi} \frac{\cos \varphi}{\rho} {\bf \hat y} \int ^\infty _{-\infty} \frac{\delta '(z)}{\sqrt{K +(z-z_0)^2}} dz \nonumber \\
 &=& \frac{gq}{2 \pi} \frac{\cos \varphi}{\rho} {\bf \hat y}\left[ \frac{\delta (z)}{\sqrt{K +(z-z_0)^2}}{\bigg \vert}_{-\infty} ^{~\infty} + \int ^\infty _{-\infty} \frac{\delta (z) (z-z_0)}{(K+(z-z_0)^2)^{3/2}}\right] \\
 &=& \frac{gq}{2 \pi} \frac{\cos \varphi}{\rho} \left(\frac{-z_0}{(K+ z_0^2)^{3/2}} \right) {\bf \hat y}~. \nonumber
\end{eqnarray}
 The surface term in the integration by parts is zero and all the non-z dependent terms are packed into the ``constant" $K=\rho ^2+ x_0^2 -2 x_0 \rho \cos \varphi$. Next the the $d \rho$ integration gives
  \begin{eqnarray}
 \label{dp}
 &-&\frac{gq z_0 \cos \varphi}{2 \pi} {\bf \hat y} \int ^\infty _0 \frac{1}{\rho (\rho ^2 +x_0^2 + z_0^2 -2 x_0 \rho \cos \varphi )^{3/2}} \rho d\rho  \nonumber \\
&=& - \frac{gq z_0 \cos \varphi}{2 \pi} {\bf \hat y} \left( \frac{2(1+ x_0 \cos \varphi /r_0 )}{x_0 ^2 + 2 z_0 ^2 - x_0^2 \cos 2 \varphi}\right) ~,
\end{eqnarray}
with $r_0 = \sqrt{x_0^2 +z_0^2}$ - recall that since we choose $y_0=0$ this effectively means $\rho_0 = x_0$ and thus $r_0 =\sqrt{\rho_0^2 + z_0^2} = \sqrt{x_0^2 +z_0^2}$. Finally we carry out the $d \varphi$ integration 
  \begin{equation}
 \label{dphi}
{\bf P}_{mech} ^{hid} = -\frac{gq z_0 }{2 \pi} {\bf \hat y} \int ^{2 \pi} _0   \left( \frac{2\cos \varphi (1+ x_0 \cos \varphi /r_0 )}{x_0 ^2 + 2 z_0 ^2 - x_0^2 \cos 2 \varphi}\right) d\varphi =  -\frac{gq z_0 }{2 \pi} {\bf \hat y} \left( \frac{-2 \pi}{x_0 r_0}\right) \to \frac{qg z_0}{\rho_0 r_0} {\bf \hat y}~,
\end{equation}
This hidden mechanical momentum accounts for and cancels the field momentum of \eqref{momentum-2} (${\bf P}_{mech} ^{hid} + {\bf P}_{EM} ^{disk} = 0$) and the center of energy theorem is saved.

One potential way to observe the presence of the disk singularity would be to move the electric charge with respect to the monopole {\it i.e.} change $x_0$ and/or $z_0$. From \eqref{dphi} this would change ${\bf P}^{hid} _{mech}$, or in other words the charges making up the hidden mechanical momentum of the disk would accelerate, and thus in general would radiate. One should be able to see this radiation and thus experimentally distinguish the set-up of monopole field plus disk singularity from a pure monopole.

\section{Banderet monopole}

There is another formulation of magnetic charge which has a sheet magnetic field similar to that of the Wu-Yang formulation from equation \eqref{b-disk}. This is known as the Banderet monopole formulation \cite{banderet} and has a vector potential of the form
\begin{equation}
    \label{band}
    {\bf A}_{band} = - \frac{g}{r} \varphi \sin \theta ~ {\bf \hat \theta}~.
\end{equation}
The vector potential ${\bf A}_{band}$ is related to the Dirac string potentials ${\bf A} _\pm$ from \eqref{A-coulomb} by the gauge transformation
\begin{equation}
    \label{band-1}
    {\bf A}_{band} = {\bf A}_\pm  - g \nabla [(\pm 1 - \cos \theta) \varphi] =  - \frac{g}{r} \varphi \sin \theta ~ {\bf \hat \theta}~.
\end{equation}
In contrast to the Dirac string potential ${\bf A}_\pm$, the vector potential ${\bf A}_{band}$ from \eqref{band} does not have a string singularity, but it is non-single valued due to the presence of $\varphi$. This non-single valuedness of ${\bf A}_{band}$ leads to a sheet magnetic field. Taking the curl of ${\bf A}_{band}$ (see equation 1.64 of \cite{shnir} for details) one finds
\begin{equation}
    \label{band-2}
    {\bf B} = \nabla  \times {\bf A}_{band} = \frac{g}{r^2} {\bf \hat r} - 2 \pi g \Theta (x) \delta (y) x \frac{{\bf \hat r}}{r^2} ~.
\end{equation}
The first term in \eqref{band-2} is the expected Coulomb magnetic field. The second term is a sheet magnetic field that is the positive, half-infinite $xz$-plane ({\it i.e.} the $xz$-plane with $x \ge 0$) that is a result of the Heaviside step function, $\Theta (x)$, and the Dirac delta function, $\delta (y)$. This is in contrast to the second term of the Wu-Yang magnetic field from \eqref{b-disk} which is the whole of the $xy$-plane. The second term in \eqref{band-2} is a mixture of spherical terms ({\it i.e.} the $\frac{{\bf \hat r}}{r^2}$ term) and Cartesian coordinate ({\it i.e.} the $\Theta (x), \delta (y)$ and $x$ terms). 

We now calculate the field momentum associated with placing an electric charge $q$ in the magnetic field ${\bf B}$ from \eqref{band-2}. To simplify the calculation we place the electric charge at $y_0 {\bf \hat y}$. The electric field is then ${\bf E} = \frac{{\bf r}'}{(r')^3}$ where ${\bf r}' = {\bf r} - y_0 {\bf \hat y}$ and $r' = (x^2 + (y-y_0)^2 +z^2)^{1/2}$. As before the field momentum associated with the Coulomb part of the magnetic field is zero 
\begin{equation}
\label{band-3}
{\bf P}_{EM} ^{Coulomb} = \frac{1}{4 \pi} \int q \frac{{\bf r}'}{(r')^3} \times g \frac{{\bf r}}{r^3} d^3 x = 0
\end{equation}
It is the second term - the half infinite sheet in \eqref{band-2} - that leads to a non-zero field momentum. 
\begin{eqnarray}
\label{band-4}
{\bf P}_{EM} ^{Sheet} &=& -\frac{2 \pi qg}{4 \pi} \int  \Theta (x) \delta (y) x  \frac{{\bf r}'}{(r')^3} \times \frac{{\bf r}}{r^3} d^3 x  \nonumber \\
&=& -\frac{qgy_0}{2}  \int _0 ^\infty dx \int _{-\infty} ^{\infty} dz
\frac{x^2 {\bf \hat z}+ x z {\bf \hat x}}{(x^2+y_0^2 +z^2)^{3/2} (x^2+ z^2)^{3/2}} \\
&=& \frac{qgy_0}{2}  \int _0 ^\infty dx \int _{-\infty} ^{\infty} dz
\frac{x^2 {\bf \hat z}}{(x^2+y_0^2 +z^2)^{3/2} (x^2+ z^2)^{3/2}}~. \nonumber
\end{eqnarray}
In the second line of \eqref{band-4} we have carried out the $dy$ integration using the $\delta (y)$, we have carried out the cross product ${\bf r}' \times {\bf r}$, and implemented the $\Theta (x)$ in the range of $dx$ integration. In the third line we have done the $dz$ integration of the ${\bf \hat x}$ component which is zero since the integrand of the ${\bf \hat x}$ component is an odd function of $z$ over an even integration range. 

Using Mathematica to carry out the remaining $dx$ and $dz$ integration in \eqref{band-4} gives the result $\int _0 ^\infty ...dx \int _{-\infty} ^{\infty} ... dz = \frac{\pi}{2 y_0^2}$ which inserting back into \eqref{band-4} gives
\begin{equation}
\label{band-5}
{\bf P}_{EM} ^{Sheet} = \frac{qg \pi}{4 y_0} {\bf \hat z}~.
\end{equation}
This example again shows that these delta function string, sheet, or half sheet contributions to the magnetic fields of monopoles are not unique the the Dirac or Wu-Yang formulations, but are a generic feature. One can also show that the above system has a hidden mechanical momentum which balances the electromagnetic field momentum of \eqref{band-5} and thus preserves the center of energy theorem.

\section{Summary and conclusions}

In this work the Wu-Yang fiber bundle approach to magnetic charge \cite{wu-yang,yang} has been extended with a disk, delta function contribution to the magnetic field \eqref{b-disk} and an associated electromagnetic field momentum \eqref{momentum-2}. This is similar to the field momentum recently found in the Dirac string model of magnetic charge \cite{siva}. In section V we have also shown that the non-singular, non-single valued Banderet vector potential for a monopole also has a delta function half disk contribution to the magnetic field, and an associated electromagnetic field momentum. These three examples indicate that these delta functions additions are generic features of Abelian monopoles.  

There are two ways to deal with this electromagnetic field momentum: (i) Allow that this uncancelled field momentum leads to a violation of the center of energy theorem \cite{zangwill,coleman}. (ii) There is some ``hidden" momentum in the system which cancels the field angular momentum. By direct calculation we have shown that there is a ``hidden" momentum, given in equation \eqref{dphi}, in the current density which exists at the interface where the two potential meet. This ``hidden" mechanical momentum exactly balances the field momentum. However, by accepting this ``hidden" mechanical momentum, one has to conclude that the Wu-Yang construction is not a pure monopole, but rather a monopole plus a current density of  ${\bf J} = -\frac{g}{2 \pi \rho}  \delta ' (z) {\hat {\bf \varphi}}$. Along with the work in \cite{siva} this shows that Abelian monopoles of the Dirac type or the Wu-Yang type are not pure monopoles, but rather monopoles plus some additional current density. This leaves only the topological 't Hooft-Polyakov monopoles \cite{thooft,polyakov}, the recently investigated electroweak topological monopoles \cite{hung1,hung2} or the two-vector potential model \cite{cabibbo,zwanziger,singleton-1996} as viable models for magnetic charge.   

In concluding we note that there are other works which point toward the reality of the Dirac string when one considers the gravitational effect of the Dirac string \cite{banyas}. Here our argument for the reality of the extra string and disk contributions is in the context of the electromagnetic interaction alone.  \\

{\bf Acknowledgment:} DS is a 2023-2024 KITP Fellow at the Kavli Institute for Theoretical Physics. This research was supported in part by the National Science Foundation under Grant No. NSF PHY-1748958. DS and SMG acknowledge useful discussions with P.Q. Hung.

\end{document}